\documentclass[prl,showpacs,floatfix,twocolumn,letterpaper,nofootinbib]{revtex4}
\usepackage{bm,amsmath}
\usepackage{dcolumn}
\usepackage{natbib}

\newcommand{\Eref}[1]{Eq.~(\ref{#1})}
\newcommand{\tref}[1]{Table~\ref{#1}}
\newcommand{\rtw}{\rightarrow}

\begin{document}

\title{Enhanced sensitivity to time-variation of $m_p/m_e$ in
the inversion spectrum of ammonia}
\author{V. V. Flambaum$^{1,2}$}
\author{M. G. Kozlov$^{3,1}$}
\affiliation{$^1$School of Physics, University of New South Wales,
Sydney, 2052 Australia}
\affiliation{$^2$Institute for Advanced Study,
Massey University (Albany Campus), Private Bag 102904, North Shore MSC Auckland, New
Zealand}
\affiliation{$^3$Petersburg Nuclear Physics Institute,
Gatchina, 188300, Russia}
\date{ \today }

\begin{abstract}
We calculate the sensitivity of the inversion spectrum of ammonia to
possible time-variation of the ratio of the proton mass to the
electron mass, $\mu=m_p/m_e$. For the inversion transition
($\lambda\approx 1.25$~cm$^{-1}$) the relative frequency shift is
significantly enhanced: $\delta \omega/\omega=-4.46\, \delta
\mu/\mu$. This enhancement allows one to increase sensitivity to the
time-variation of $\mu$ using NH$_3$ spectra for high redshift
objects. We use published data on microwave spectra of the object
B0218+357 to place the limit  $\delta \mu/\mu =(0.6\pm 1.9)\times
10^{-6}$ at redshift $z=0.6847$; this limit is  several times better
than the limits obtained by different methods and may be
significantly improved. Assuming linear time dependence we obtain
$\dot{\mu}/\mu=(-1 \pm 3) \times 10^{-16}$~yr$^{-1}$.
\end{abstract}

\maketitle

\section{Introduction}\label{intro}

The possible time-variation of the fundamental constants has been
discussed for a long time. The interest in this discussion has grown
considerably after the recent discovery of the acceleration of the
universe. The latter is usually regarded as evidence for the
existence of dark energy. Cosmological evolution of dark energy may
cause variations in fundamental constants, such as the
fine-structure constant $\alpha$ and the proton to electron mass
ratio, $\mu\equiv m_p/m_e$.
The electron mass is one of the parameters of the Standard Model, it
is proportional to the vacuum expectation value of the Higgs field
(the weak scale). The proton mass is proportional to another
fundamental parameter, the quantum chromodynamics scale
$\Lambda_{QCD}$ ($m_p \approx 3\Lambda_{QCD}$). The proportionality
coefficients cancel out in the relative variation. Therefore, we are
speaking about the relative variation of a very important
dimensionless fundamental parameter of the Standard Model, the ratio
of the strong to weak scale, defined as
 $\delta(\Lambda_{QCD}/m_e)/(\Lambda_{QCD}/m_e)=\delta\mu/\mu $.

It is known that $\mu$ defines the scales of electronic,
vibrational, and rotational intervals in molecular spectra,
$E_\mathrm{el}:E_\mathrm{vib}:E_\mathrm{rot}\sim
1:\mu^{-1/2}:\mu^{-1}$. Similarly, the ratio of electronic and
hyperfine intervals in atoms and molecules also depend on $\mu$,
$E_\mathrm{el}\!:\!E_\mathrm{hfs}\!\sim\!1\!:\!\alpha^2 g_p
\mu^{-1}$, where $g_p$ is the proton $g$-factor. These scalings are
used to look for the time-variation of $\mu$ by comparing
electronic, vibrational, rotational, and hyperfine spectra of atoms
and molecules \cite{VP96,DWB98}. In the most recent astrophysical
studies \cite{RBH06} a non-zero effect was reported for two quasars
at $3.5\,\sigma$ level:
\begin{equation}\label{mu_var1}
    \delta \mu/\mu =(20\pm 6)\times 10^{-6},
\end{equation}
at a time scale of approximately 12 Gyr. Assuming linear variation
with time this result translates into $\dot{\mu}/\mu=(-17\pm 5)
\times 10^{-16}$~yr$^{-1}$. A different method, comparison of the
hyperfine transition in atomic hydrogen with optical transitions in
ions, was used in Refs.~\cite{TWM05,TWM07}. This method allows one
to study variation of the parameter $x=\alpha^2g_p/\mu$. Analysis of
9 quasar spectra with redshifts $0.23 \le z \le 2.35$ gave
\begin{align}\label{x_var}
    \delta x/x &=(6.3\pm 9.9)\times 10^{-6},\\
\label{x_vara}
     \dot{x}/x &=(-6\pm 12)\times 10^{-16}~\mathrm{yr}^{-1},
\end{align}
which is consistent with zero variation of $\mu$. In
Refs.~\cite{CK03,KCG04,KCL05} the 18~cm $\lambda$-doublet lines in
OH molecule were studied from objects at the redshifts $z\approx
0.247$, $z\approx 0.6847$, and $z\approx 0.765$ and no
time-variation of the parameter $g_p(\alpha^2\mu)^{\nu}$ was seen,
where $\nu \lesssim 2$.


In this paper we use the enhanced sensitivity of the inversion
spectrum of ammonia to variation of $\mu$ to place a new limit on
the time-variation of $\mu$ at the cosmological timescale. The
NH$_3$ molecule has a pyramidal shape and the inversion frequency
depends on the exponentially small tunneling of three hydrogens
through the potential barrier \cite{TS55}. Because of that it is
very sensitive to any changes of the parameters of the system,
particularly to the reduced mass for this vibrational
mode~\cite{VKB04}.

We use high-resolution ammonia spectra for gravitational lens
B0218+357, published by \citet{HJK05}. The redshifts for ammonia
lines are compared to the redshifts for the rotational lines of
other molecules measured in Refs.~\cite{HJK05,WC95,CW95}. The
ammonia lines have an order of magnitude stronger dependence on
$\mu$ than the usual vibrational lines; this enhancement allows us
to place the best limit on the variation of $\mu$.

\section{Inversion spectrum of {NH}$_3$}\label{spectrum}

The inversion spectrum of NH$_3$ has been studied for a very long
time \cite{TS55} and is considered a classical example of the
tunneling phenomenon. The inversion vibrational mode is described by
a double well potential with first two vibrational levels lying
below the barrier. Because of the tunneling, these two levels are
split in inversion doublets. The lower doublet corresponds to the
wavelength $\lambda\approx 1.25$~cm and is used in ammonia masers.
Molecular rotation leads to the centrifugal distortion of the
potential curve. Because of that the inversion splitting depends on
the rotational angular momentum $J$ and its projection on the
molecular symmetry axis $K$:
 \begin{align}\label{w_inv}
 \omega_\mathrm{inv}(J,K) = \omega^0_\mathrm{inv}
 - c_1
 \left[J(J+1)-K^2\right] + c_2 K^2\,,
 \end{align}
where we omitted terms with higher powers of $J$ and $K$.
Numerically, $\omega^0_\mathrm{inv}\approx 23.787$~GHz, $c_1\approx
151.3$~MHz, and $c_2\approx 59.7$~MHz.

In addition to the rotational structure \eqref{w_inv} the inversion
spectrum includes much smaller hyperfine structure. For the main
nitrogen isotope $^{14}$N, the hyperfine structure is dominated by
the electric quadrupole interaction ($\sim 1$~MHz)~\cite{HT83}.
Because of the dipole selection rule $\Delta K=0$ the levels with
$J=K$ are metastable and in laboratory experiments the width of the
corresponding inversion lines is usually determined by collisional
broadening. In astrophysics, the hyperfine structure for spectra
with high redshifts is not resolved and we will not discuss it here.

For our purposes it is important to know how the parameters in
\eqref{w_inv} depend on fundamental constants. One can measure only
dimensionless ratios of frequencies which do not depend on the units
used. It is convenient to consider all parameters in atomic units.
The energy unit Hartree is $E_H=m_e e^4/\hbar ^2=e^2/a_B$, where
$a_B$ is the Bohr radius ($E_H$=2 Ry=219475~cm$^{-1}$). In these
units all electron energies ($E_e/E_H$) and electrostatic potentials
($U(r)/E_H$) have no dependence on the fundamental constants (here
we neglect small relativistic corrections which give a weak $\alpha$
dependence), the vibrational intervals $\sim \mu^{-1/2}$ and the
rotational intervals $\sim \mu^{-1}$. The inversion frequency
$\omega^0_\mathrm{inv}/E_H$ and constants $c_{1,2}/E_H$ are also
functions of $\mu$ only (see below). Note that the coefficients
$c_i$ depend on $\mu$ through the reduced mass of the inversion mode
and because they are inversely proportional to the molecular moments
of inertia. That implies a different scaling of
$\omega^0_\mathrm{inv}$ and $c_i$ with $\mu$. The magnetic hyperfine
structure of NH$_3$ is due to the interaction of nuclear magnetic
moments and proportional to $\alpha^2g_p^2\mu^{-2}$.

We see that different frequencies in the inversion spectrum scale
differently with $\mu$ and $\alpha$. In principle, this allows one to
study time-variation of $\mu$ and $\alpha$ by comparing different
lines of the inversion spectrum. On the other hand, it may be
 preferable to use independent references (see below).

\section{Inversion Hamiltonian}\label{H_inv}

The inversion spectrum \eqref{w_inv} can be approximately described
by the following Hamiltonian:
\begin{align}\label{H_inv1}
H_\mathrm{inv} &= -\tfrac{1}{2M_1} \partial^2_x+U(x)\\
&+\tfrac{1}{I_1(x)}\left[J(J+1)-K^2\right]
+\tfrac{1}{I_2(x)}K^2,
\nonumber
\end{align}
where $x$ is the distance from N to the H-plane, $I_1$, $I_2$ are
moments of inertia perpendicular and parallel to the molecular axis
correspondingly and $M_1$ is the reduced mass for the inversion
mode. If we assume that the length $d$ of the N---H bond does not
change during inversion, then $M_1=2.54m_p$ and
\begin{align}\label{moment_I1}
 I_1(x) &\approx \tfrac{3}{2}m_p d^2 \left[1+0.2(x/d)^2\right], \\
 \label{moment_I2}
 I_2(x) &\approx 3m_p d^2 \left[1-(x/d)^2\right].
\end{align}
The dependence of $I_{1,2}$ on $x$ generates a correction to the
potential energy of the form $C(J,K)x^2/\mu$. This changes the
vibrational frequency and the effective height of the potential
barrier, therefore changing the inversion frequency
$\omega_\mathrm{inv}$ given by \Eref{w_inv}.

Following \cite{SI62} we can write the potential $U(x)$ in
\eqref{H_inv1} in the following form:
\begin{align}\label{H_inv2}
U(x) &= \tfrac{1}{2}k x^2 +b \exp\left(-c x^2\right).
\end{align}
Fitting vibrational frequencies for NH$_3$ and ND$_3$ gives
$k\approx 0.7598$~a.u., $b\approx 0.05684$~a.u., and $c\approx
1.3696$~a.u. Numerical integration of the Schr\"{o}dinger equation
with potential \eqref{H_inv2} gives the following result:
\begin{align}
 \label{dw_inv6}
 \frac{\delta\omega_\mathrm{inv}}{\omega_\mathrm{inv}}
 &\approx -4.46\, \frac{\delta\mu}{\mu}\,.
\end{align}
It is instructive to reproduce this result from an analytical
calculation. In the semiclassical approximation the inversion
frequency is estimated as \cite{LL77}:
\begin{subequations}
\begin{align}
 \label{w_inv1}
 \omega_\mathrm{inv}
 &= \frac{\omega_v}{\pi}\exp\left(-S\right) \\
 \label{w_inv2}
 &= \frac{\omega_v}{\pi}
 \exp\left(-\frac{1}{\hbar}\int_{-a}^a \sqrt{2M_1(U(x)-E)}
 \mathrm{d} x\right),
\end{align}
\end{subequations}
where $\omega_v$ is the vibrational frequency of the inversion mode,
$S$ is the action in units of $\hbar$, $x=\pm a$ are classical
turning points for the energy $E$. For the lowest vibrational state
$E=U_\mathrm{min}+\tfrac12 \omega_v$. Using the experimental values
$\omega_v$=950 cm$^{-1}$ and $\omega_\mathrm{inv}$=0.8 cm$^{-1}$, we
get $S\!\approx\! 5.9\,$.

Expression \eqref{w_inv2} allows one to calculate the dependence of
$\omega_\mathrm{inv}^0$ on the mass ratio $\mu$. Let us present $S$
in the following form: $S = A\mu^{1/2}\int_{-a}^{a}
\sqrt{(U(x)-E)/E_H}d(x/a_B)$, where $A$ is a numerical constant. We
see that the dependence of $\omega_\mathrm{inv}^0$ on $\mu$ appears
from the factor $\mu^{1/2}$ in $S$ and from the vibrational
frequency $\omega_v$ and $E-U_\mathrm{min}=\tfrac12 \omega_v$ which
are proportional to $\mu^{-1/2}$. Below we assume that all energies
are measured in atomic units and omit the atomic energy unit $E_H$.
Then we obtain
\begin{subequations}
\begin{align}
 \label{dw_inv1}
 \frac{\mathrm{d}\omega_\mathrm{inv}^0}{\mathrm{d}\mu}
 &= -\omega_\mathrm{inv}^0
 \left(\frac{1}{2\mu}+
 \frac{\mathrm{d}S}{\mathrm{d}\mu} \right)\\
 \label{dw_inv2}
 &= -\omega_\mathrm{inv}^0
 \left(\frac{1}{2\mu}+
 \frac{\partial S}{\partial\mu}
 +\frac{\partial S}{\partial E}\frac{\partial E}{\partial\mu}
 \right),
\end{align}
\end{subequations}
where we took into account that $\partial S/\partial a =0$ because
the integrand in \eqref{w_inv2} turns to zero at $x=\pm a$.

It is easy to see that ${\partial S}/{\partial\mu}=S/2\mu$. The
value of the third term in \Eref{dw_inv2} depends on the form of the
potential barrier:
\begin{align}
 \label{dw_inv3}
 \frac{\partial S}{\partial E}
 &=-\frac{q}{4}\frac{S}{U_\mathrm{max}-E},
\end{align}
where for the square barrier $q=1$, and for the triangular barrier
$q=3$. For a more realistic barrier shape $q\approx 2$. Using
parametrization \eqref{H_inv2} to determine $U_\mathrm{max}$ we get:
\begin{align}
 \label{dw_inv4}
 \!\frac{\delta\omega_\mathrm{inv}^0}{\omega_\mathrm{inv}^0}
 &\approx -\frac{\delta\mu}{2\mu}
 \left(1+S
 +\frac{S}{2}\frac{\omega_v}{U_\mathrm{max}-E}
 \right)
 = -4.4\, \frac{\delta\mu}{\mu}.
\end{align}

We see that the inversion frequency of NH$_3$ is an order of
magnitude  more sensitive to the change of $\mu$ than typical
vibrational frequencies. The reason for that is clear from
\Eref{dw_inv4}: it is the large value of the action $S$ for the
tunneling process.

Let us also find the dependence of the constants $c_{1,2}$ on $\mu$
in \Eref{w_inv}. According to Eqs.~\eqref{H_inv1}
--~\eqref{moment_I2} both constants must have the same dependence on
$\mu$. Below we focus on the constant $c_2$, which is linked to the
last term in Hamiltonian \eqref{H_inv1}. It follows from
\Eref{moment_I2} that this term generates a correction to the
potential:
\begin{align}
 \label{dw_rot1}
 \delta U(x) &= \frac{K^2}{3m_pd^4} x^2.
\end{align}
This correction does not change the height of the barrier, but
changes the energy $E=U_\mathrm{min}+\tfrac12 \omega_v$ in
\eqref{w_inv2} by raising the potential minimum and increasing the
vibrational frequency:
\begin{align}
 \label{dw_rot2}
 U_\mathrm{min} &\rightarrow U_\mathrm{min}
 + \frac{K^2}{3m_pd^4} x_0^2,\\
 \label{dw_rot3}
 \omega_v &\rightarrow \omega_v
 \left(1 + \frac{K^2}{3m_pd^4 k} \right).
\end{align}
With the help of \Eref{dw_inv3} with $q=2$ we can find the constant
$c_2$:
\begin{align}
 \label{dw_rot4}
 c_2 = \frac{\omega^0_\mathrm{inv}}{3m_pd^4 k}
 \left(1+\frac{k x_0^2+\omega_v}{U_\mathrm{max}-E}S\right).
\end{align}
We can differentiate \Eref{dw_rot4} to estimate how $c_2$ depends on
$\mu$. This leads to: $\delta c_2/c_2=-5.0\,\delta\mu/\mu$, while
the numerical solution with Hamiltonian \eqref{H_inv1} gives:
\begin{align}
 \label{dw_rot5}
 \frac{\delta c_{1,2}}{c_{1,2}}
 &= -5.1\frac{\delta\mu}{\mu}\,.
\end{align}

It is clear that NH$_3$ is not the only molecule with enhanced
sensitivity to variation of $\mu$. Similar enhancement should take
place for all tunneling transitions in molecular spectra. For
example, the inversion frequency for ND$_3$ molecule is 15 times
smaller than for NH$_3$ and \Eref{w_inv1} leads to $S\approx 8.4$,
compared to $S\approx 5.9$ for NH$_3$. According to \Eref{dw_inv4}
that leads to a slightly higher sensitivity of the inversion
frequency to $\mu$ \cite{VKB04}:
\begin{align}
 \label{dw_nd3}
 \mathrm{ND_3:}
 \left\{
 \begin{array}{rcl}
   \frac{\delta\omega_\mathrm{inv}}{\omega_\mathrm{inv}}
   &\approx &-5.7\, \frac{\delta\mu}{\mu}\,,\\
\\
   \frac{\delta c_2}{c_2}
   &\approx &-6.2\,\frac{\delta\mu}{\mu}\,.
 \end{array}
 \right.
\end{align}

\section{Redshifts for molecular lines in the microwave spectra of B0218+357}
\label{redshifts}

In the previous section we saw that the inversion frequency
$\omega_\mathrm{inv}^0$ and the rotational intervals
$\omega_\mathrm{inv}(J_1,K_1)-\omega_\mathrm{inv}(J_2,K_2)$ have
different dependencies on the constant $\mu$. In principle, that
allows one to study time-variation of $\mu$ by comparing different
intervals in the inversion spectrum of ammonia. For example, if we
compare the rotational interval to the inversion frequency, then
Eqs.~\eqref{dw_inv6} and~\eqref{dw_rot5} give:
\begin{align}
 \label{red1}
 \frac{\delta\{[\omega_\mathrm{inv}(J_1,K_1)-\omega_\mathrm{inv}(J_2,K_2)]
 /\omega^0_\mathrm{inv}\}}
 {[\omega_\mathrm{inv}(J_1,K_1)-\omega_\mathrm{inv}(J_2,K_2)]/\omega^0_\mathrm{inv}}
 &= - 0.6 \frac{\delta\mu}{\mu}\,.
\end{align}
The relative effects are substantially larger if we compare the inversion
transitions with the  transitions between the quadrupole and magnetic
 hyperfine components. However,
in practice this method will not work because of the smallness of
the hyperfine structure compared to  typical line widths in
astrophysics.

It is more promising to compare the inversion spectrum of NH$_3$ with
rotational spectra of other molecules, where
\begin{align}
 \label{red2}
 \frac{\delta\omega_\mathrm{rot}}{\omega_\mathrm{rot}}
 &= - \frac{\delta\mu}{\mu}\,.
\end{align}
In astrophysics any frequency shift is related to a corresponding
redshift:
\begin{align}
 \label{red3}
 \frac{\delta\omega}{\omega}
 &= - \frac{\delta z}{1+z}\,.
\end{align}
According to Eqs.~\eqref{dw_inv6} and~\eqref{red2}, for a given
astrophysical object with $z=z_0$ variation of $\mu$ will lead to a
change of the redshifts of all rotational lines $\delta
z_\mathrm{rot}=(1+z_0)\,{\delta\mu}/{\mu}$ and corresponding shifts
of all inversion lines of ammonia $\delta z_\mathrm{inv}=
4.46\,(1+z_0)\,{\delta\mu}/{\mu}$. Therefore, comparing the redshift
for NH$_3$ with the redshifts for rotational lines we can find
${\delta\mu}/{\mu}$:
\begin{align}
 \label{red4}
 \frac{\delta\mu}{\mu}
 &= 0.289\, \frac{z_\mathrm{inv}-z_\mathrm{rot}}{1+z_0}\,.
\end{align}

\begin{table}[tbh]
  \caption{Redshifts for molecular rotational lines, ammonia inversion lines,
  and hydrogen hyperfine line in the spectrum of B0218+357.}
  \label{tab1}
\begin{tabular}{lrllc}
\hline \hline
\multicolumn{5}{c}{Rotational lines}\\
CO       &$  J=1\rtw 2  $&   red-shifted    & 0.68470     &  \cite{WC95}\\
         &$             $&   blue-shifted   & 0.68463     &  \cite{WC95}\\
\multicolumn{2}{l}{CO, HCO$^+$, HCN}
                         & average          & 0.68466(1)  &  \cite{CW97}\\
\hline
\multicolumn{5}{c}{Inversion lines of NH$_3$}\\
 NH$_3$  &$(J,K)=(1,1)  $&   red-shifted    & 0.684679(3) &  \cite{HJK05}\\
         &$             $&  blue-shifted    & 0.684649(15)&  \cite{HJK05}\\
         &$     =(2,2)  $&   red-shifted    & 0.684677(3) &  \cite{HJK05}\\
         &$             $&  blue-shifted    & 0.684650(17)&  \cite{HJK05}\\
         &$     =(3,3)  $&   red-shifted    & 0.684673(3) &  \cite{HJK05}\\
         &$             $&  blue-shifted    & 0.684627(33)&  \cite{HJK05}\\
&\multicolumn{2}{c}{average red-shifted}
                                    & 0.684676(3)\\
&\multicolumn{2}{c}{average blue-shifted}
                                    & 0.684647(11)\\
\hline
H        &$\lambda=21$~cm& average          &  0.68466(4) &  \cite{CRY93}\\
\hline\hline
\end{tabular}
\end{table}

In \tref{tab1} we list the redshifts for microwave lines in the
spectrum of the object B0218+357. Three inversion lines
$(J,K)=(1,1),\,(2,2),\,\mathrm{and}\,(3,3)$ are reported in
Ref.~\cite{HJK05}. Each of them consists of a narrow red-shifted and
a wide blue-shifted component. The splitting between the red-shifted
and blue-shifted components, which is about 5 km/s, is ascribed to
the complicated structure of the molecular cloud \cite{HJK05}. Using
average redshifts of these inversion components
 (0.684676(3) and 0.684647(11)) from \tref{tab1} we can
calculate the average deviation of the inversion redshift in respect
to the average molecular redshift (0.68466(1)):
\begin{align}
 \label{red5a}
 \Delta z_\mathrm{av}^\mathrm{unweighted} &= (0.2 \pm 0.9) \times 10^{-5}\,,\\
 \label{red5b}
 \Delta z_\mathrm{av}^\mathrm{weighted} &= (0.6 \pm 0.9) \times
 10^{-5}\,.
\end{align}
\Eref{red4} gives the following estimate for variation of $\mu$:
\begin{align}
 \label{red6}
 \frac{\delta\mu}{\mu}
 &=
 10^{-6}\times \left\{
 \begin{array}{ll}
  0.3 \pm 1.6 & \mathrm{(unweighted)},\\
  1.1 \pm 1.5 & \mathrm{(weighted)}.\\
\end{array}
 \right.
\end{align}
As a final result we present a conservative limit
with  larger error bars to cover the total interval between the minimal
and maximal values  for both estimates:
\begin{equation} \label{final}
 \frac{\delta\mu}{\mu}=( 0.6 \pm 1.9)\times 10^{-6}.
\end{equation}

We can also compare averaged redshift for ammonia with that of
hydrogen to get a restriction on the variation of the parameter
$y=\alpha^2g_p\mu^{3.46}$:
\begin{align}
 \label{red7}
 \frac{\delta y}{y}
 &=
 \frac{z_\mathrm{inv}-z_\mathrm{hfs}}{1+z_0}
 = (1 \pm 17) \times 10^{-6}\,.
\end{align}

The estimates (\ref{red6}--\ref{red7}) can be further improved by
dedicated analysis of the molecular spectra published in
Refs.~\cite{WC95,CW95,HJK05}. As it was mentioned in \cite{WC95},
the majority of molecular lines from B0218+357 have two velocity
components. The same applies to the hydrogenic 21~cm line
\cite{MWF01}. Instead of taking an average, as we have done in
(\ref{red6}--\ref{red7}), all red-shifted and all blue-shifted
components should be analyzed independently. That may allow one to
reduce the error bars significantly.

We thank M.~Kuchiev for helpful discussions and J.~Ginges for
reading the manuscript. This work is supported by the Australian
Research Council, Godfrey fund and Russian foundation for Basic
Research, grant No. 05-02-16914.


\end{document}